\documentclass[11pt]{article}
\usepackage[dvips]{color}
\usepackage[dvips]{xcolor}
\usepackage[dvips]{graphicx}
\usepackage{times}
\usepackage{amssymb}
\usepackage{amsmath}
\usepackage{stmaryrd}
\usepackage{amscd}
\usepackage{latexsym}
\usepackage{bbm}

\catcode`@=11
\renewcommand{\section}{\@startsection{section}{1}{0pt}{\medskipamount}
{\medskipamount}{\large\bf}}
\numberwithin{equation}{section}
\catcode`@=12

\newcommand{\cv}{}

\def\a{\alpha}
\def\b{\beta}

\def\d{\delta}

\def\g{\gamma}
\def\p{\psi}

\def\la{\lambda}

\def\s{\sigma}
\def\t{\tau}

\def\sfrac#1#2{{\textstyle\frac{#1}{#2}}}

\def\ic{{\mathrm i}}

\def\pa{\partial}
\def\>{\rangle}
\def\<{\langle}
\def\+{\dagger}
\def\={\ =\ }
\def\ax{\a(x)}
\def\bx{\b(x)}
\def\gx{\g(x)}
\def\hb{{\hbar}}
\def\bt{\bar{\tau}}
\def\uh{U_{\text{hom}}}

\newcommand{\R}{\mathbb R}

\newcommand{\cN}{{\cal N}}
\newcommand{\unity}{\mathbbm{1}}
\newcommand{\be}{\begin{equation}}
\newcommand{\ee}{\end{equation}}
\newcommand{\beq}{\begin{equation*}}
\newcommand{\eeq}{\end{equation*}}
\newcommand{\bea}{\begin{eqnarray}}
\newcommand{\eea}{\end{eqnarray}}
\newcommand{\bal}{\begin{aligned}}
\newcommand{\eal}{\end{aligned}}

\newcommand{\und}{\qquad{\textrm{and}}\qquad}

\begin{document}

\begin{titlepage}
\setcounter{page}{0}
\begin{flushright}
ITP--UH--21/08\\
\end{flushright}

\vskip 2.0cm

\begin{center}

{\Large\bf 
N=4 Mechanics, WDVV Equations and Polytopes \footnote{
Talk at the XVII International Colloquium on Integrable Systems and
Quantum Symmetries in Prague, 19-21 June 2008, and \\ \phantom{.}\quad\ \
at the XXVII International Colloquium on Group Theoretical Methods in Physics
in Yerevan, 13-19 August 2008}
}
\vspace{12mm}

{\large 
Olaf Lechtenfeld 
} 
\\[8mm]
\noindent {\em 
Institut f\"ur Theoretische Physik, Leibniz Universit\"at Hannover} 

\vspace{12mm}

\begin{abstract}
\noindent
$N{=}4$ superconformal $n$-particle quantum mechanics on the 
real line is governed by two prepotentials, $U$ and $F$, 
which obey a system of partial nonlinear differential 
equations generalizing the Witten-Dijkgraaf-Verlinde-Verlinde 
(WDVV) equation for $F$. The solutions are encoded by the 
finite Coxeter systems and certain deformations thereof, 
which can be encoded by particular polytopes.
We provide $A_n$ and $B_3$ examples in some detail.
Turning on the prepotential~$U$ in a given $F$~background
is very constrained for more than three particles and nonzero 
central charge. The standard ansatz for~$U$ is shown to fail
for all finite Coxeter systems. Three-particle models are 
more flexible and based on the dihedral root systems.
\end{abstract}

\end{center}

\end{titlepage}

\section{Conformal quantum mechanics: Calogero system}

\noindent
We are investigating systems of $n{+}1$ identical point particles with 
unit mass whose motion on the real line is governed by the Hamiltonian
\be\label{h}
H\=\sfrac{1}{2} p_i p_i\ +\ V_B (x^1, \dots, x^{n+1}) 
\ee
and subject to the canonical quantization relations
\be
[x^i, p_j]\=\ic {\d_j}^i \ .
\ee
Together with
\be
D\=-\sfrac{1}{4} (x^i p_i +p_i x^i) \und
K\= \sfrac{1}{2} x^i x^i \ ,
\ee
this Hamiltonian realizes an $so(1,2)$ conformal algebra
\be\label{al} 
[D,H]\=-\ic H\ ,\quad
[H,K]\=2\ic D\ ,\quad
[D,K]\=\ic K 
\ee
if the potential is homogeneous of degree~$-2$,
\be
(x^i\pa_i+2)\,V_B\=0 \ .
\ee
When demanding also permutation and translation invariance
as well as admitting only two-body forces, the solution is uniquely
given by the Calogero potential,
\be
V_B \= \sum_{i<j} \sfrac{g^2}{(x^i-x^j)^2} \ .
\ee

\section{$\cN{=}4$ superconformal extension: $su(1,1|2)$ algebra}

\noindent
Let us extend the algebra from $so(1,2)\simeq su(1,1)$ 
to the superalgebra $su(1,1|2)$ with central charge~$C$ 
by enlarging the set of generators
\begin{equation*}
(H,D,K)\ \to\ (H,D,K,Q_\a,S_\a,J_a,C) 
\qquad\textrm{with}\quad \a=1,2\ ,\quad a=1,2,3 \ ,\quad 
{(Q_\a)}^\+{=}{\bar Q}^\a\ ,\quad {(S_\a)}^\+{=}{\bar S}^\a
\end{equation*}
and imposing the nonvanishing (anti)commutators:
\begin{align*}
&
[D,H]=-\,\ic H\  && [H,K]=2\ic D\ 
\\[4pt]
&
[D,K]=+\ic K\  && [J_a,J_b]=\ic \epsilon_{abc} J_c\ 
\\[4pt]
&
\{ Q_\a, \bar Q^\b \}=2 H {\d_\a}^\b\  &&
\{ Q_\a, \bar S^\b \}=
+2\ic {{(\s_a)}_\a}^\b J_a-2 D {\d_\a}^\b-\,\ic C {\d_\a}^\b\ 
\\[4pt]
&
\{ S_\a, \bar S^\b \}\ =2 K {\d_\a}^\b\  &&
\{ \bar Q^\a, S_\b \}=
-\,2\ic {{(\s_a)}_\b}^\a J_a-2 D {\d_\b}^\a+\ic C {\d_\b}^\a\ 
\\
& [D,Q_\a] = -\sfrac{\ic}{2} Q_\a\  && 
[D,S_\a] =+\sfrac{\ic}{2} S_\a\ 
\\[4pt]
&
[K,Q_\a] =+\ic S_\a\  && [H,S_\a] =-\ic Q_\a\ 
\\[4pt]
&
[J_a,Q_\a] =-\sfrac{1}{2} {{(\s_a)}_\a}^\b Q_\b \qquad\qquad && 
[J_a,S_\a] =-\sfrac{1}{2} {{(\s_a)}_\a}^\b S_\b\ 
\\[4pt]
& [D,\bar Q^\a] =-\sfrac{\ic}{2} \bar Q^\a\  && 
[D,\bar S^\a] =+\sfrac{\ic}{2} \bar S^\a\ 
\\[4pt]
& [K,\bar Q^\a] =+\ic \bar S^\a\  && 
[H,\bar S^\a] =-\ic \bar Q^\a\ 
\\[4pt]
&
[J_a,\bar Q^\a] =\sfrac{1}{2} \bar Q^\b {{(\s_a)}_\b}^\a\  && 
[J_a,\bar S^\a] =\sfrac{1}{2}
\bar S^\b {{(\s_a)}_\b}^\a \quad.
\end{align*}

To realize this algebra one must pair the bosonic coordinates~$x^i$
with fermionic partners~$\psi^i_\a$ \ and \ $\bar\psi^{i\a}{=}{\psi^i_\a}^\+$
\ with \ $i=1,\dots,n{+}1$ \ and \ $\a=1,2$ \ subject to
\be 
\{\p^i_\a, \p^j_\b \}=0\ , \qquad \{ {\bar\p}^{i\a}, {\bar\p}^{j\b} \}=0\ ,
\qquad \{\p^i_\a, {\bar\p}^{j\b} \}\= {\d_\a}^\b \d^{ij} \ .
\ee
Surprisingly, the non-interacting generator candidates
\be\label{QSfree} 
{Q_0}_\a\=p_i \p^i_\a\ , \qquad 
\bar Q_0^\a\=p_i \bar\p^{i\a} \und
{S_0}_\a\=x^i \p^i_\a\ , \qquad
\bar S_0^\a\=x^i \bar\p^{i\a}\ ,
\ee
\be\label{HDKJfree} 
H_0= \sfrac12 p_i p_i\ ,\quad
D_0= -\sfrac{1}{4}(x^i p_i +p_i x^i)\ ,\quad
K_0= \sfrac12 x^i x^i\ ,\quad
{J_0}_a = \sfrac{1}{2} \bar\p^{i\a} {{(\s_a)}_\a}^\b \p^i_\b 
\ee
fail to obey the $su(1,1|2)$ algebra, and hence interactions are needed!
Their simplest implementation changes only
\be\label{Qcorr} 
Q_\a \= Q_{0\a} -\ic\,[S_{0\a},{\cv V}] \und
H \= H_0 + {\cv V} \ ,
\ee
just requiring the invention of a potential~$V(x,\p,\bar\p)$.

A minimal ansatz to close the $su(1,1|2)$ algebra reads~\cite{Wyl99,BGL04}
\be\label{ans}  
V \= V_B(x)\ -\
U_{ij}(x) \langle \p^i_\a {\bar\p}^{j\a} \rangle\ +\
\sfrac14 F_{ijkl}(x) \langle\p^i_\a\p^{j\a}\bar\p^{k\b}\bar\p^l_\b\rangle 
\ee
where $\langle\ldots\rangle$ denotes symmetric (Weyl) ordering.
The coefficient functions $U_{ij}$ and $F_{ijkl}$ are totally symmetric 
and homogeneous of degree~$-2$. With this, the supersymmetry generators
in~(\ref{Qcorr}) become
\be\label{Qform}
Q_\a \= \bigl(p_j-\ic\,x^i\,U_{ij}(x)\bigr)\,\p_\a^j \ \,-
\sfrac{\ic}{2}\,x^i\,F_{ijkl}(x)\,\<\p^j_\b\,\p^{k\b}\bar\p^l_\a\> \ .
\ee

\section{The structure equations: WDVV, flatness, homogeneity}

\noindent
Inserting the minimal $V$~ansatz~(\ref{ans}) into the $su(1,1|2)$ algebra 
and demanding its closure produces conditions on $U_{ij}$ and $F_{ijkl}$.
First, one learns that
\be\label{pot} 
U_{ij} \= \pa_i\pa_j U \und F_{ijkl} \= \pa_i\pa_j\pa_k\pa_l F \ ,
\ee
introducing two scalar prepotentials $U$ and $F$. Second, these prepotentials 
are subject to the ``structure equations''~\cite{Wyl99,BGL04}
\bea
&&  
(\pa_i\pa_k\pa_p F)(\pa_p\pa_l\pa_j F)\=
(\pa_i\pa_l\pa_p F)(\pa_p\pa_k\pa_j F) 
\quad,\qquad 
x^i \partial_i \partial_j \partial_k F\=-\d_{jk} 
\label{w1} \\[6pt]
&& 
\ \pa_i\pa_j U -(\pa_i\pa_j\pa_k F)\,\pa_k U\=0 
\quad,\qquad\qquad\qquad\qquad\qquad\ \ 
x^i \pa_i U\=-C \ .
\label{w2}
\eea
The quadratic equation for~$F$ is the famous WDVV~equation~\cite{W,DVV}.
The relation below it (linear in~$U$) resembles a covariant constancy
equation, and we label it as the ``flatness condition''. Its consistency
implies the WDVV equation contracted with~$\pa_i U$. Both the WDVV~equation
and the flatness condition trivialize when contracted with~$x^i$.
Finally, the two right equations are homogeneity properties for $F$ and~$U$.
One of their consequences is
\be
x^i\,F_{ijkl} \= -\pa_j\pa_k\pa_lF \und x^i\,U_{ij} \= -\pa_jU \ .
\ee
The one for~$F$ may be integrated twice to
\be
(x^i\pa_i - 2) F \= -\sfrac12\,x^ix^i \ .
\ee
Clearly, there is the redundancy of adding a quadratic polynomial to~$F$
and a constant to~$U$. The third outcome of the $su(1,1|2)$ algebra is
\be
V_B \= \sfrac12\,(\pa_iU)(\pa_iU) \ +\
\sfrac{\hb^2}{8}\,(\pa_i\pa_j\pa_kF)(\pa_i\pa_j\pa_kF) \ ,
\ee
where we have reinstalled~$\hbar$ to exhibit the quantum part in~$V_B$.

In case of vanishing central charge, $C{=}0$, a partial solution
consists in putting \ $U\equiv0$.
Since $U$ does not enter in~(\ref{w1}), the natural strategy is to
firstly solve the WDVV equation and secondly
turn on a flat~$U$ in this $F$~background.

\section{Prepotential ansatz: covectors and couplings}

\noindent
The homogeneity conditions \ 
$(x^i\pa_i-2)F=-\sfrac12\,x^ix^i$ \ und \ $x^i\pa_i U=-C$ \
are solved by~\cite{Wyl99}
\be\label{covans}
F \= -\sfrac12\sum_{\a} f_\a\ \ax^2\,\ln|\ax| \ +\ F_{\textrm{hom}}
\und U \= -\sum_{\a} g_\a\,\ln|\ax|\ +\ U_{\textrm{hom}}\ ,
\ee
where $F_{\textrm{hom}}$ and $U_{\textrm{hom}}$ are arbitrary homogeneous 
functions of degree~$-2$ and~$0$, respectively. The sums run over
a set of real covectors~$\a$ (not indexed!) with values \ $\ax\=\a_ix^i$,
which are subject to the constraints
\be\label{hcond}
\sum_\a\,f_\a\ \ax^2 \= x^i x^i\ =:\ R^2 \und
\sum_\a\, g_\a \= C \ .
\ee
The coefficients $f_\a$ are essentially fixed by~(\ref{hcond}) and
(if positive) may be absorbed into a rescaling of~$\a$, while the $g_\a$
will emerge as coupling constants which, however, may be frozen to zero.
One may rewrite the expressions~(\ref{covans}) as
\be\label{radans}
F \= -\sfrac12\,R^2\ln R \ +\ F'_{\textrm{hom}} \und
U \= -C\,\ln R \ +\ U'_{\textrm{hom}}\ ,
\ee
or linearly combine (\ref{covans}) and~(\ref{radans}) with coefficents
adding to one.

Due to the generality of~$F_{\textrm{hom}}$, we are currently unable to
solve the WDVV~equation~(\ref{w1}) with (\ref{covans}) or~(\ref{radans}),
except for $F^{(\prime)}_{\textrm{hom}}{\equiv}0$.
Even then, the nonlinearity of~(\ref{w1}) restricts the linear combinations to
\be\label{ansatz}
F \= -\sfrac12\smash{\sum_{\a}} f_\a\ \ax^2\,\ln|\ax| \qquad\textrm{or}\qquad
F \= +\sfrac12\smash{\sum_{\a}} f_\a\ \ax^2\,\ln|\ax|\ -\ R^2\ln R\ ,
\ee
and imposes~\cite{MarGra99,Ves99}
\be \label{w3}
\sum_{\a,\b} f_\a f_\b\,
\frac{\a{\cdot}\b}{\ax\,\bx}\,(\a\wedge\b)^{\otimes2}\=0 
\qquad\textrm{with}\qquad
(\a\wedge\b)^{\otimes2}_{ijkl} \= (\a_i\b_j-\a_j\b_i)(\a_k\b_l-\a_l\b_k)\ .
\ee

Thus, let us limit ourselves to the ansatz~(\ref{ansatz}) and try to turn
on~$U$. Even this is too difficult in general, so let us drop the homogeneous
pieces in (\ref{covans}) and~(\ref{radans}) and just combine the inhomogeneous
parts. Then, the flatness condition~(\ref{w2}) rules out all `$R$' terms
in $F$ or~$U$ and demands
\be \label{w4}
\sum_{\b} \Bigl( g_\b\,\frac{1}{\bx}\ -\ f_\b \sum_\a g_\a\,
\frac{\a{\cdot}\b}{\ax} \Bigr)\,\frac{1}{\bx}\ \b\otimes\b \=0
\qquad\textrm{with}\qquad
(\b\otimes\b)_{ij} \= \b_i \b_j \ ,
\ee
while the bosonic potential reads
\be 
V_B \= \sfrac12 \sum_{\a,\b} \frac{\a{\cdot}\b}{\ax\;\bx}\,
\Bigl( g_\a g_\b\ +\ \sfrac{\hb^2}{4} f_\a f_\b\,(\a{\cdot}\b)^2 \Bigr)\ .
\ee

Because the equations decouple for mutually orthogonal sets of covectors,
it suffices to take $\{\a\}$ as being indecomposable. 
In particular, it is convenient in translation-invariant models
to decouple the center of mass \ $\a_{\textrm{com}}(x)=\sum_i x^i$, 
reducing the bosonic configuration space from $\R^{n+1}$ to $\R^n$. 
Note that this alters 
\be
R^2 \= x^ix^i \qquad\textrm{to}\qquad
R^2 \= \bigl(x^i-\sfrac{1}{n+1}\a_{\textrm{com}}(x)\bigr)^2 \=
\sfrac{1}{n+1}\bigl[ n\textstyle{\sum_i} x^i x^i - 
2\textstyle{\sum_{j<\ell}} x^k x^\ell \bigr]\ .
\ee
Partial results are known for $n{\le}3$
\cite{Wyl99,ChaVes01,BGL04,GLP07,FeiVes07,GLP08},
but the case $n{=}2$ is special since the WDVV~equation is empty then,
which admits many extra solutions.

\section{$U{=}0$ solutions: root systems}

\noindent
It was shown by Martini and Gragert~\cite{MarGra99} and extended by
Veselov~\cite{Ves99} that the set~$\Phi^+$ of positive roots of 
any simple Lie algebra solves the left equation in~(\ref{w3}).
Let us normalize the long and short roots as
\be
\a{\cdot}\a=2 \quad\textrm{for}\quad \a\in\Phi^+_{\textrm{L}} \und 
\b{\cdot}\b=\sfrac{2}{r} \quad\textrm{for}\quad \b\in\Phi^+_{\textrm{S}}\ ,
\qquad\textrm{with}\quad r=1,2,3\ .
\ee
Recalling that
\be \label{recall}
\sum_{\a\in\smash{\Phi^+}} \a\otimes\a \= h^\vee\,\unity \und
\sum_{\a\in\smash{\Phi^+}} \sfrac{2}{\a{\cdot}\a}\,\a\otimes\a \= h\,\unity
\ee
are determined by the Coxeter number~$h$ and the dual Coxeter number~$h^\vee$,
the left condition in~(\ref{hcond}) becomes
\be\label{fnorm}
\unity \= \sum_\a f_\a\,\a\otimes\a \=
f_{\textrm{L}} \sum_{\a\in\smash{\Phi^+_{\textrm{L}}}} \a\otimes\a \ +\
f_{\textrm{S}} \sum_{\a\in\smash{\Phi^+_{\textrm{S}}}} \a\otimes\a \=
f_{\textrm{L}}\,\sfrac{r\,h^\vee\!-h}{r-1}\,\unity \ +\ 
f_{\textrm{S}}\,\sfrac{h-h^\vee}{r-1}\,\unity\ ,
\ee
which is solved by the one-parameter family
\be \label{ffamily} 
f_{\textrm{L}} \= \sfrac{1}{h^\vee} + (\sfrac{h\ }{h^\vee}{-}1)\,t \und
f_{\textrm{S}} \= \sfrac{1}{h^\vee} - (r{-}\sfrac{h\ }{h^\vee})\,t 
\qquad\textrm{for}\quad t\in\R\ .
\ee

\begin{figure}[ht]
\centerline{\includegraphics[width=10cm]{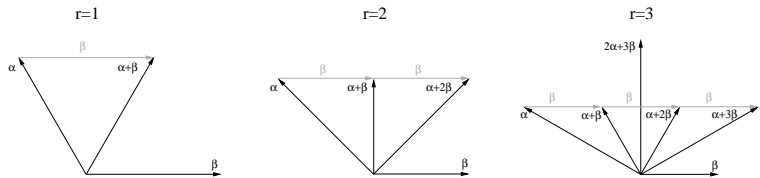}}
\caption{Short-root strings through a long root}
\label{fig:1}
\end{figure}
It is not hard to see that the subset of roots belonging to any plane
spanned by a short root~$\b$ and its string 
$(\a,\a{+}\b,\a{+}2\b,\ldots,\a{+}r\b)$ through a long root~$\a$ makes
the double sum in the left equation of~(\ref{w3}) already vanish.
Since the full double sum decomposes into contributions of such planes,
we get a prepotential solution
\be\label{Fsol}
\begin{aligned}
F(t)&\= -\sfrac12\, \Bigl(
f_{\textrm{L}}\!\! \sum_{\a\in\smash{\Phi^+_{\textrm{L}}}} \ + \
f_{\textrm{S}}\!\! \sum_{\a\in\smash{\Phi^+_{\textrm{S}}}} \Bigr)\,
\ax^2\,\ln|\ax| \\[4pt]
&\= -\sfrac{1}{2h^\vee} \sum_{\a\in\smash{\Phi^+}} \ax^2\,\ln|\ax| \ - \ 
\sfrac{t}{2}\,\Bigl[ 
(\sfrac{h\ }{h^\vee}{-}1)\!\!\sum_{\a\in\smash{\Phi^+_{\textrm{L}}}} -\,
(r{-}\sfrac{h\ }{h^\vee})\!\!\sum_{\a\in\smash{\Phi^+_{\textrm{S}}}} \Bigr]\,
\ax^2\,\ln|\ax|
\end{aligned}
\ee
which is unique only for simply-laced Lie algebras.
Note that $|f_{\textrm{L}}|$ and $|f_{\textrm{S}}|$ may be absorbed into
a rescaling of~$\a$ but their signs cannot, and so the non-simply-laced 
solution generalizes the $t{=}0$ one found before~\cite{MarGra99,Ves99}
by adding to it a concrete~$F_{\textrm{hom}}$.

Let us give two examples, with $n{+}1$ and 3 particles, respectively: 
\bea
A_n\oplus A_1:&&
\{\ax\} \= \bigl\{\ 
x^i{-}x^j \ ,\ \textstyle{\sum_i} x^i 
\quad\big|\quad 1\le i<j\le n{+}1\ \bigr\}
\und f_\a \= \sfrac{1}{n+1} \ ,\\[4pt] \nonumber
G_2\oplus A_1:&& 
\{\ax\} \= \bigl\{\ 
\sfrac{1}{\sqrt{3}}(x^i{+}x^{j}{-}2x^{k})\ ,\
\sfrac{1}{\sqrt{3}}(x^i{-}x^{j})\ ,\ x^1{+}x^2{+}x^3 
\quad\big|\quad (i,j,k)\quad\textrm{cyclic}\ \bigr\} \\ &&
\textrm{and}\qquad 
f_{\textrm{L}}\=\sfrac14+\sfrac12t\ ,\quad 
f_{\textrm{S}}\=\sfrac14-\sfrac32t\ ,\quad
f_{\textrm{com}}\=\sfrac13\ .
\eea

The Weyl groups of the simple Lie algebras can be extended by the
non-crystallographic Coxeter groups $H_4$ (60 positive roots), 
$H_3$ (15 positive roots) and $I_2(p)$ ($p$ positive roots),
which also clear the WDVV equation~\cite{Ves99}.
The dihedral groups~$I_2(p)$ with $f_\a=\sfrac1p$ cover all rank-two 
root systems, including $A_1{\oplus}A_1$, $A_2$, $BC_2$ and $G_2$ for
$p=2,3,4$ and $6$, respectively, upon rescaling of~$\a$.
\begin{figure}[ht]
\centerline{\includegraphics[width=10cm]{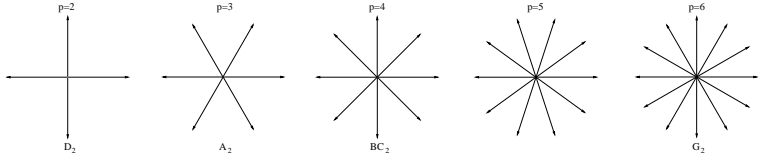}}
\caption{Root systems of the dihedral groups $I_2(p)$ for $p=2,3,4,5,6$.}
\label{fig:2}
\end{figure}

\section{$U{=}0$ solutions: deformed root systems}

\noindent
The Lie-algebra root systems are only the tip of an iceberg of WDVV solutions.
It has been shown~\cite{ChaVes01,FeiVes07} that certain deformations of them
retain the WDVV property. Let us rephrase some examples in our terminology.

The three positive roots of $A_2$ may be rearranged as the edges of 
an equilateral triangle. Consider now a deformation of this triangle,
keeping the incidence relation \ $\a+\b-\g=0$.
The homogeneity condition~(\ref{hcond}) (and therefore also the WDVV~equation)
is easily solved by \ $f_\a \= \frac{|\b\cdot\g|}{4\,A^2}$ \ and cyclic
permutations, where $A$ denotes the area of the triangle.

If we try the same idea on the $A_3$ system, we obtain the six edges of
a regular tetrahedron and deform to encounter the five-dimensional
moduli space of tetrahedral shapes (modulo scale). 
\begin{figure}[ht]
\centerline{\includegraphics[width=3cm]{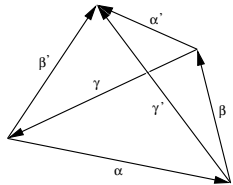}}
\caption{Tetrahedral configuration of covectors}
\label{fig:3}
\end{figure}
Again, the homogeneity condition~(\ref{hcond}) has a unique solution~$f_\a$,
but now the WDVV equation enforces the three conditions
\be \label{ortho}
\a\cdot\a' =0\ ,\quad \b\cdot\b' =0\ ,\quad \g\cdot\g' =0
\ee
on the skew edge pairs.
These relations restrict the above moduli space to the three-dimensional 
subspace of orthocentric tetrahedra (modulo scale), with
\be
f_\a = \frac{|\b{\cdot}\g\ \b'{\cdot}\g'|}{36\,V^2} \und
f_{\a'} = \frac{|\b{\cdot}\g'\ \b'{\cdot}\g|}{36\,V^2} \und \textrm{cyclic}\ ,
\ee
where $V$ is the volume.
Alternatively, we may implement the conditions~(\ref{ortho}) by picking
three non-coplanar covectors, say $\a'$, $\b'$ and $\g'$, scaling them
such that 
\be
\a'\cdot\b' \= \b'\cdot\g' \= \g'\cdot\a' \= 1
\ee
and employing the three-dimensional vector product in
fixing the remaining three covectors via
\be
\a \= \a'\times(\b'\times\g') \= \b'-\g'\ ,\quad
\b \= \b'\times(\g'\times\a') \= \g'-\a'\ ,\quad
\g \= \g'\times(\a'\times\b') \= \a'-\b'\ .
\ee
With these data one gets \ $6\,V=\a'{\cdot}(\b'{\times}\g')$ \ as well as
\be
f_\a = \frac{\a'{\cdot}\a'{-}1}{36\,V^2} \und
f_{\a'} = \frac{(\b'{\cdot}\b'{-}1)(\g'{\cdot}\g'{-}1)}{36\,V^2} 
\und \textrm{cyclic}\ .
\ee

In fact, this strategy generalizes to orthocentric $n$-simplices as 
$n$-parametric deformations of the regular $n$-simplex generated by
the $\sfrac12n(n{+}1)$ positive roots of~$A_n$, with
\be
f_\a \= \frac{|\b{\cdot}\g\ \b'{\cdot}\g'\ \b''{\cdot}\g''\
\cdots\ \b^{(n-2)}{\cdot}\g^{(n-2)}|}{(n!\ V)^2} \quad\textrm{etc.}\ .
\ee
The orthocentricity derives from the WDVV~equation by a the following
dimensional reduction argument. Take \ $\hat{n}_i x^i\to\infty$ \
for some fixed covector~$\hat{n}$. Then, any factor $\frac1\ax$ in the
WDVV~equation~(\ref{w3}) vanishes unless \ $\a{\cdot}\hat{n}=0$,
which amounts to a reduction of the covector set~$\{\a\}$ to its intersection
with the hyperplane orthogonal to~$\hat{n}$. This process may be iterated 
until only covectors laying in a plane $\a{\wedge}\b$ spanned by two covectors
$\a$ and~$\b$ survive. This situation admits two possibilities:
either the $\a$ and $\b$ are concurrent, in which case another covector
$\a{+}\b$ or $\a{-}\b$ completes a triangle satisfying the WDVV equation,
or else $\a$ and $\b$ are skew, in which case there is no further covector
in their plane and WDVV demands orthogonality.

\begin{figure}[ht]
\centerline{\includegraphics[width=3cm]{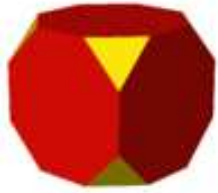}}
\caption{Truncated cube}
\label{fig:4} 
\end{figure}
The $B_3$ root system provides another example.
Four copies of the 3 short and 6 long positive roots can be assembled
into the edge set of a truncated cube. We deform this polyhedron to
\be
\bigl\{\a(x)\bigr\} \= \bigr\{
d_1x^1\ ,\ d_2x^2\ ,\  d_3x^3\ ;\  
c_3(c_2x^1{\pm}c_1x^2)\ ,\  c_1(c_3x^2{\pm}c_2x^3)\ ,\  c_2(c_1x^3{\pm}c_3x^1)
\bigr\}
\ee
with $c_i, d_i\in\R$, retaining the `incidence relations' 
of a truncated cuboid. For \ $c^2:=c_0^2+c_1^2+c_2^2+c_3^2$ \ and
\be
\bigl\{ f_\a \bigr\} \= \Bigl\{
\sfrac{c_0^2+c_1^2-c_2^2-c_3^2}{c^2\ d_1^2}\ ,\
\sfrac{c_0^2-c_1^2+c_2^2-c_3^2}{c^2\ d_1^2}\ ,\
\sfrac{c_0^2-c_1^2-c_2^2+c_3^2}{c^2\ d_1^2}\ ;\
\sfrac{1}{c^2\,c_3^2}\ ,\ \sfrac{1}{c^2\,c_1^2}\ ,\ \sfrac{1}{c^2\,c_2^2}
\Bigr\}
\ee
we satisfy the homogeneity condition~(\ref{hcond}), 
i.e.~$\sum_\a f_\a\,\a{\otimes}\a=\unity$.
The relevant combinations $\sqrt{f_\a}\,\a$ depend only on the three ratios
$c_i/c_0$. The rigid $B_3$ root system with~(\ref{ffamily}) occurs for \
$c_1=c_2=c_3=1$ \ and \ $c_0^2=\frac{2-3t}{1+t}$.
The case of~$C_3$ is very similar.

Finally, let us present an example based on weights rather than roots,
namely a deformation of the $B_3$ representation 
$\underline{7}\oplus\underline{8}$, i.e.~the vector plus spinor weights.
For the 3 positive `vector' and 4 positive `spinor' covectors we take
\be
\ax=d_1 x^1\ ,\quad \bx=d_2 x^2\ ,\quad \gx=d_3 x^3\ ;\quad
\sfrac{\a{+}\b{+}\g}{2}\ ,\quad \sfrac{\a{-}\b{-}\g}{2}\ ,\quad
\sfrac{-\a{+}\b{-}\g}{2}\ ,\quad \sfrac{-\a{-}\b{+}\g}{2}
\ee
with $d_i\in\R$, keeping the relations between vector and spinor weights. 
For \ $d^2:=d_1^2+d_2^2+d_3^2$ \ and
\be
f_\a=\sfrac{-d_1^2+d_2^2+d_3^2}{d^2\,d_1^2}\ ,\quad
f_\b=\sfrac{ d_1^2-d_2^2+d_3^2}{d^2\,d_2^2}\ ,\quad
f_\g=\sfrac{ d_1^2+d_2^2-d_3^2}{d^2\,d_3^2}\quad\textrm{and}\quad
f_{\textrm{spinor}}= \sfrac{2}{d^2}
\ee
we obey~(\ref{hcond}) and achieve a two-parameter deformation of the
original weight system at \ $c_1=c_2=c_3$. The corresponding polyhedron,
whose edges are built from 4 copies of the vector and 6 copies of the
spinor weights, is a (inhomogeneously scaled) rhombic dodecahedron
with the faces dissected into triangles.
\begin{figure}[ht]
\centerline{\includegraphics[width=3cm]{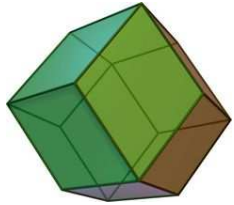}}
\caption{Rhombic dodecahedron}
\label{fig:5} 
\end{figure}

It is important to realize that all examples fulfil the WDVV equation,
because the above dimensional reduction argument applies. The crucial
properties are the mutual orthogonality of non-concurrent non-parallel
edges as well as the incidence relations, which `sew' the triangles
together into a polyhedron. Yet, these properties are only necessary
but not sufficient. Finally we remark that all our examples are
part of a larger moduli space of $n{=}3$ families of WDVV solutions
\cite{ChaVes01,FeiVes07}.

\section{$U{\neq}0$ solutions: no-go `theorem' for $n{>}2$}

\noindent
Recall that, for turning on
\be \label{upart}
U \= -\sum_{\a} g_\a\,\ln|\ax|
\qquad\textrm{with}\qquad \sum_\a g_\a \= C
\ee
in a given $F$ background determined by~$\{\a,f_\a\}$,
we need to solve the flatness condition~(\ref{w4}).
In principle, we may modify~(\ref{w4}) by adding a homogeneous term 
$U_{\textrm{hom}}$ to the prepotential above, but let us postpone
this option for the time being. Then,
matching the coefficients of the double poles in~(\ref{w4}) requires that
\be \label{UF3}
\textrm{either}\qquad g_\b = 0 \qquad\textrm{or else}\qquad 
\b{\cdot}\b\ f_\b = 1 \qquad\textrm{for each covector $\b$}\ .
\ee

In the undeformed irreducible root-system solutions, the Weyl group identifies
the $f_\a$ and $g_\a$ coefficients for all roots of the same length. Hence,
besides the $f_{\textrm{L}}$ and $f_{\textrm{S}}$ values in~(\ref{ffamily})
we have couplings $g_{\textrm{L}}$ and $g_{\textrm{S}}$ for a number
$p_{\textrm{L}}$ and $p_{\textrm{S}}$ of long and short positive roots,
respectively.\footnote{
For expliciteness, $p_{\textrm{L}}=\sfrac{n}{2}\,\sfrac{rh^\vee-h}{r-1}$
and $p_{\textrm{S}}=\sfrac{n}{2}\,\sfrac{r(h-h^\vee)}{r-1}$, with the
sum $p=p_{\textrm{L}}+p_{\textrm{S}}=\sfrac{n}{2}h$.}
This simplifies the trace of~(\ref{fnorm}) to
\be \label{rootsumrule}
n\= \sum_{\a} \a{\cdot}\a\,f_\a \=
2f_{\textrm{L}}\,p_{\textrm{L}}\ +\
\sfrac2r f_{\textrm{S}}\,p_{\textrm{S}}
\qquad\longrightarrow\qquad p_{\textrm{L}}+p_{\textrm{S}} \= n 
\qquad\textrm{if}\quad g_{\textrm{L}},g_{\textrm{S}}\neq0\ .
\ee
Since the total number~$p_{\textrm{L}}+p_{\textrm{S}}$ of positive roots 
always exceeds~$n$ (except for $A_1^{\oplus n}$), we are forced
to put either $g_{\textrm{S}}=0$ or $g_{\textrm{L}}=0$.
Therefore, all simply-laced root systems are ruled out!
For the $r{>}1$ root systems, we get
\bea \label{case1}
&\textrm{either} \qquad &
g_{\textrm{S}}=0\ ,\quad g_{\textrm{L}}=g
\qquad\buildrel{(\ref{UF3})(\ref{rootsumrule})}\over\longrightarrow\qquad
f_{\textrm{S}}=\sfrac{r}{2}\sfrac{n-p_{\textrm{L}}}{p_{\textrm{S}}}\le0\ ,\quad
f_{\textrm{L}}=\sfrac12 \\[6pt] \label{case2}
&\textrm{or}\phantom{nnl} \qquad &
g_{\textrm{S}}=g\ ,\quad g_{\textrm{L}}=0
\qquad\buildrel{(\ref{UF3})(\ref{rootsumrule})}\over\longrightarrow\qquad
f_{\textrm{S}}=\sfrac{r}{2}\ ,\quad
f_{\textrm{L}}=\sfrac12\sfrac{n-p_{\textrm{S}}}{p_{\textrm{L}}}\le0\ .
\eea
We see that in the non-simply-laced one-parameter family (\ref{ffamily})
there is always one member which obeys (\ref{case1}) or~(\ref{case2}).
For it, we must still check the remainder of~(\ref{w4}),
\be \label{w5}
\sum_{\a\neq\b} g_\a f_\b\,\frac{\a{\cdot}\b}{\ax\ \bx}\ \b\otimes\b \= 0\ .
\ee
Even though its trace is alway satisfied, the traceless part is violated
for any nontrivial root system with the data (\ref{case1}) or~(\ref{case2}).
Hence, there do not exist $U$~solutions of the standard form~(\ref{upart}) 
for any Coxeter root system. Perhaps this no-go result may be overcome 
by adding suitable $U_{\textrm{hom}}$ contributions. Certainly it can
be avoided for $n{=}2$ because in this case (\ref{ansatz}) may be relaxed
(see below). Finally, we have not yet studied the flatness conditions for 
the deformed root systems of the previous section.

\section{$U{\neq}0$ solutions: dihedral solutions for $n{=}2$}

\noindent
As mentioned before, the case of three particles with translation invariance,
i.e.~$n{=}2$, is special for the absence of the WDVV~equation. 
In fact, it is easy to see that any set~$\{\a\}$ of covectors can be made
to obey the left condition in~(\ref{hcond}) with suitably chosen~$f_\a$. 
To study concrete examples, we look at the most symmetric cases, 
namely the dihedral root systems mentioned earlier.

It is crucial that we take advantage of the freedom at $n{=}2$ to add
`radial terms' in our ansatz:
\bea \label{Fdih}
F \!\!&=&\!\! -\sfrac12\sum_{\a} f_\a\ \ax^2 \ln|\ax| -\sfrac12 f_R\,R^2\ln R  
\quad\longrightarrow\quad
\sum_\a f_\a\,\a{\otimes}\a \= (1{-}f_R)\,\unity \ ,\\[4pt] \label{Udih}
U \!\!&=&\!\! -\sum_{\a} g_\a\,\ln|\ax| -g_R\,\ln R 
\qquad\qquad\qquad\longrightarrow\quad
\sum_\a g_\a \= C{-}g_R \ .
\eea
The flatness condition then reduces to (\ref{UF3}) and the trace of~(\ref{w5}) 
plus the relation \ $g_R+(C{-}g_R)f_R=0$. 
It is obeyed for the $I_2(p)$ system if \
$(g_{\textrm{even}},g_{\textrm{odd}}){=:}(g_{\textrm{S}},g_{\textrm{L}})$ \
when $p$ is~even and if \ $g_\a{=:}g\ \forall\a$ \ when $p$~is odd.
Turning on $g$~couplings for all covectors fixes 
$\a{\cdot}\a\,f_\a=1$, and so we obtain
\be
p=2\,(1{-}f_R) \qquad\longrightarrow\qquad
g_R=\sfrac{p-2}{p}\,C \qquad\longrightarrow\qquad
\sfrac2pC=\sum_\a g_\a=\Bigl\{\begin{smallmatrix} 
\sfrac{p}{2}(g_{\textrm{S}}{+}g_{\textrm{L}}) \ & \textrm{for $p$ even} \\
p\,g & \textrm{for $p$ odd} \end{smallmatrix}\ \ .
\ee

In order to ease the interpretation as three-particle systems,
we embed the relative-motion configuration space~$\R^2$ into
$\R^3\ni(x^1,x^2,x^3)$ \ and rotate such that \ $\a_{\textrm{com}}=(1,1,1)$.
For identical particles we require invariance under permutations \
$(x^1,x^2,x^3)\to(x^{\pi_1},x^{\pi_2},x^{\pi_3})$ \ of the full three-body 
coordinates. This limits $p$ to multiples of~3. The `radial coordinate'
then becomes
\be
R^2 \= \sfrac13\bigl\{(x^{12})^2+(x^{23})^2+(x^{31})^2\bigr\}
\= \sfrac23\bigl\{(x^1)^2+(x^2)^2+(x^3)^2-x^1x^2-x^2x^3-x^3x^1\bigr\}\ .
\ee
\begin{figure}[ht]
\centerline{\includegraphics[width=4cm]{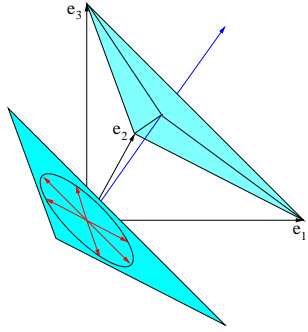}}
\caption{Embedding of $A_2$ roots into $\R^3$}
\label{fig:6}
\end{figure}

\section{Examples}

\noindent
For illustration we explicitly display the $I_2(p)$ solutions based on
(\ref{Fdih}) and (\ref{Udih}) for the first few values of~$p$.

$\underline{p{=}2}\ :\quad A_1{\oplus}A_1$ model \hfill
$f_R{=}0\,,\quad g_R{=}0\quad\longrightarrow\quad
g_{\textrm{S}}{+}g_{\textrm{L}}{=}C$
\beq
\frac{\ax}{\sqrt{\a{\cdot}\a}}\ \in\ \Bigl\{
\sfrac1{\sqrt{2}}(x^1{-}x^2)\,,\, 
\sfrac1{\sqrt{6}}(x^1{+}x^2{-}2x^3) \Bigr\}
\qquad\qquad\qquad\qquad\qquad\qquad\qquad\ \ 
\eeq 
\beq 
V_B \ =\ \frac{g_{\textrm{S}}^2{+}\sfrac{\hb^2}{4}}{(x^1{-}x^2)^2}\ +\
\frac{3\,(g_{\textrm{L}}^2{+}\sfrac{\hb^2}{4})}{(x^1{+}x^2{-}2x^3)^2}
\qquad\qquad\qquad\qquad\qquad\qquad\qquad\qquad
\eeq

$\underline{p{=}3}\ :\quad A_2$ model \hfill
$f_R{=}{-}\sfrac12\,,\quad g_R{=}\sfrac13C\quad\longrightarrow\quad
g{=}\sfrac29C$
\beq 
\frac{\ax}{\sqrt{\a{\cdot}\a}}\ \in\ \Bigl\{
\sfrac1{\sqrt{2}}(x^1{-}x^2)\,,\,
\sfrac1{\sqrt{2}}(x^1{-}x^3)\,,\,
\sfrac1{\sqrt{2}}(x^2{-}x^3) \Bigr\} 
\qquad\qquad\qquad\qquad\qquad\quad\ \
\eeq
\beq 
V_B \= \bigl(g^2{+}\sfrac{\hb^2}{4}\bigr)\,\biggl(
\frac{1}{(x^1{-}x^2)^2}+\frac{1}{(x^2{-}x^3)^2}+\frac{1}{(x^3{-}x^1)^2}\biggr)
\ +\ \sfrac58\,\bigl(9g^2{-}\hb^2\bigr)\,\frac{1}{R^2} 
\eeq

$\underline{p{=}4}\ :\quad BC_2$ model \hfill
$f_R{=}{-}1\,,\quad g_R{=}\sfrac12C\quad\longrightarrow\quad
g_{\textrm{S}}{+}g_{\textrm{L}}{=}\sfrac14C$
\beq 
\frac{\ax}{\sqrt{\a{\cdot}\a}} \in \Bigl\{
\sfrac{x^1{-}x^2}{\sqrt{2}}\,,\, 
\sfrac{x^1{+}x^2{-}2x^3}{\sqrt{6}}\,,\,
\sfrac{\t x^1{-}\bt x^2{-}x^3}{\sqrt{3}}\,,\,
\sfrac{-\bt x^1{+}\t x^2{-}x^3}{\sqrt{3}} \Bigr\}
\quad\textrm{with}\quad \biggl\{\begin{smallmatrix}
\t = \sfrac12 (\sqrt{3}{+}1) \\[4pt] \bt = \sfrac12 (\sqrt{3}{-}1)
\end{smallmatrix}
\eeq
\beq \qquad
V_B \= \frac{g_{\textrm{S}}^2{+}\sfrac{\hb^2}{4}}{(x^1{-}x^2)^2} +
\frac{3\,(g_{\textrm{S}}^2{+}\sfrac{\hb^2}{4})}{(x^1{+}x^2{-}2x^3)^2} +
\frac{\frac32\,(g_{\textrm{L}}^2{+}\frac{\hb^2}{4})}
     {(\t x^1{-}\bt x^2{-}x^3)^2} +
\frac{\frac32\,(g_{\textrm{L}}^2{+}\frac{\hb^2}{4})}
     {({-}\bt x^1{+}\t x^2{-}x^3)^2} 
+ \frac{6(g_{\textrm{S}}{+}g_{\textrm{L}})^2{-}\frac{3}{2}\hb^2}{R^2}
\eeq

$\underline{p{=}6}\ :\quad G_2$ model \hfill
$f_R{=}{-}2\,,\quad g_R{=}\sfrac23C\quad\longrightarrow\quad
g_{\textrm{S}}{+}g_{\textrm{L}}{=}\sfrac19C$
\beq 
\frac{\ax}{\sqrt{\a{\cdot}\a}}\ \ \in\ \ \Bigl\{
\sfrac{x^1{-}x^2}{\sqrt{2}}\,,\,
\sfrac{x^1{-}x^3}{\sqrt{2}}\,,\,
\sfrac{x^2{-}x^3}{\sqrt{2}}\,,\,
\sfrac{2x^1{-}x^2{-}x^3}{\sqrt{6}}\,,\,
\sfrac{x^1{+}x^2{-}2x^3}{\sqrt{6}}\,,\,
\sfrac{-x^1{+}2x^2{-}x^3}{\sqrt{6}} \Bigr\} 
\qquad\quad
\eeq
\beq
V_B \= \frac{g_{\textrm{S}}^2{+}\sfrac{\hb^2}{4}}{(x^1{-}x^2)^2}\ +\
\frac{3\,(g_{\textrm{L}}^2{+}\sfrac{\hb^2}{4})}{(x^1{+}x^2{-}2x^3)^2}\ +\
\textrm{cyclic}\cv\ +\ 
\frac{36(g_{\textrm{S}}{+}g_{\textrm{L}})^2{-}4\hb^2}{R^2} 
\qquad\quad
\eeq

$\underline{p{=}12}\ :\quad I_2(12)$ model \hfill
$f_R{=}{-}5\,,\quad g_R{=}\sfrac56C\quad\longrightarrow\quad
g_{\textrm{S}}{+}g_{\textrm{L}}{=}\sfrac{1}{36}C$
\beq \qquad\quad\
\begin{aligned}
\frac{\ax}{\sqrt{\a{\cdot}\a}}\ \ \in\ \ \Bigl\{
\sfrac{x^1{-}x^2}{\sqrt{2}}\,,\,
\sfrac{x^1{-}x^3}{\sqrt{2}}\,,\,
\sfrac{x^2{-}x^3}{\sqrt{2}}\,,\,
\sfrac{2x^1{-}x^2{-}x^3}{\sqrt{6}}\,,\,
\sfrac{x^1{+}x^2{-}2x^3}{\sqrt{6}}\,,\,
\sfrac{-x^1{+}2x^2{-}x^3}{\sqrt{6}}\,, 
\qquad\qquad\qquad\quad\!\\[4pt] 
\sfrac{\t x^1{-}x^2{-}\bt x^3}{\sqrt{3}}\,,\,
\sfrac{\t x^1{-}\bt x^2{-}x^3}{\sqrt{3}}\,,\,
\sfrac{x^1{+}\bt x^2{-}\t x^3}{\sqrt{3}}\,,\,
\sfrac{\bt x^1{+}x^2{-}\t x^3}{\sqrt{3}}\,,\,
\sfrac{-\bt x^1{+}\t x^2{-}x^3}{\sqrt{3}}\,,\,
\sfrac{-x^1{+}\t x^2{-}\bt x^3}{\sqrt{3}} \Bigr\}
\end{aligned}
\eeq
\beq \qquad\qquad\quad
\begin{aligned}
V_B &\= \frac{g_{\textrm{S}}^2{+}\sfrac{\hb^2}{4}}{(x^1{-}x^2)^2}\ +\
\frac{3\,(g_{\textrm{S}}^2{+}\sfrac{\hb^2}{4})}{(x^1{+}x^2{-}2x^3)^2}\ +\
\frac{\frac32\,(g_{\textrm{L}}^2{+}\sfrac{\hb^2}{4})}
     {(\t x^1{-}x^2{-}\bt x^3)^2}\ +\
\frac{\frac32\,(g_{\textrm{L}}^2{+}\sfrac{\hb^2}{4})}
     {(\t x^1{-}\bt x^2{-}x^3)^2}\ +\ \textrm{cyclic} \\[4pt]
&\qquad +\ 
\frac{630(g_{\textrm{S}}{+}g_{\textrm{L}})^2{-}\frac{35}{2}\hb^2}{R^2}
\end{aligned}
\eeq

Finally, let us investigate the effect of adding to $U$ a homogeneous
piece~$\uh$ for obtaining $U_{\textrm{tot}}=U+\uh$. At $n{=}2$, all we
have to solve is the trace of the flatness condition,
\be
\pa{\cdot}\pa\,\uh\ +\
\sum_\a {\cv f_\a\,\a{\cdot}\a}\ \sfrac{\a\cdot\pa}{\ax}\,\uh\=0 
\qquad\textrm{besides}\qquad x^i\pa_i\uh=0\ .
\ee
It is convenient to pass to polar coordinates on~$\R^2$ via \
$(x^1,x^2)=(R\cos\phi,R\sin\phi)$. In the dihedral class $I_2(p)$,
the sum over the roots can be performed, and the flatness conditions
for~$U$ and for~$\uh$ is solved by
\be 
\begin{aligned}
U(R,\phi) &\= -C\,\ln R \ -\ \begin{cases}
g\,\ln|\cos(p\phi)| & \textrm{for $p$ odd} \\[4pt]
g_{\textrm{S}}\ln|\cos(\sfrac{p}{2}\phi)| + 
g_{\textrm{L}}\ln|\sin(\sfrac{p}{2}\phi)| & \textrm{for $p$ even}
\end{cases} \ \ ,\\[4pt]
\uh(\phi) &\= \sfrac{\la}{p}\,\ln|\tan(\sfrac{p}2\phi{+}\delta)|
\qquad\qquad\ \ \,\textrm{with}\qquad
\delta\=\Bigl\{ \begin{matrix} \sfrac\pi4 & \textrm{for $p$ odd} \\
0 & \textrm{\ for $p$ even} \end{matrix} \ \ \ .
\end{aligned}
\ee
After lifting to the full configuration space~$\R^3$ 
as in Figure~\ref{fig:6}, we arrive at
\be
\pa_i U_{\textrm{tot}} \= -\sum_\a g_\a\,\frac{\a_i}{\ax}\
-\ \sfrac{p-2}{p}\,C\,\frac{x_i}{R^2} \ +\ 
\la\,\biggl(\begin{smallmatrix} 
x^2{-}x^3 \\ x^3{-}x^1 \\ x^1{-}x^2 \end{smallmatrix} \biggr)\,
R^{p-2}\,\prod_\a \bigl( \ax \bigr)^{-1}\ .
\ee
For the $A_2$ model as the simplest example, one gets
\be 
F \= -\sfrac14\,\bigl( 
(x^{12})^2\ln|x^{12}|+(x^{23})^2\ln|x^{23}|+(x^{31})^2\ln|x^{31}| \bigr)
\ +\ \sfrac14\,R^2\ln R\ ,
\ee
\be
\vec{\nabla}U_{\textrm{tot}} \ \=\
\bigl[ x^{12}x^{23}x^{31} \bigr]^{-1}\,\biggl(\begin{smallmatrix}
[ \la R - g(x^{31}{-}x^{12}) ]\,x^{23} \\
[ \la R - g(x^{12}{-}x^{23}) ]\,x^{31} \\
[ \la R - g(x^{23}{-}x^{31}) ]\,x^{12} \end{smallmatrix} \biggr)
\ -\ \sfrac32\,g\,R^{-2}\,\biggl(\begin{smallmatrix}
x^1 \\ x^2 \\ x^3 \end{smallmatrix} \biggr) \ ,
\qquad\quad
\ee
\be 
\begin{aligned}
V_B^{\textrm{tot}} &\=
\bigl(g^2+\sfrac23\la^2+\sfrac{\hb^2}{4} \bigr)\,
\Bigl( \frac{1}{(x^{12})^2}+\frac{1}{(x^{23})^2}+\frac{1}{(x^{31})^2} \Bigr)
\\[4pt] & \quad
+\ \ \sfrac58 \bigl(9 g^2-\hb^2\bigr)\,R^{-2}\ +\
\la\,g\,R\,\frac{(x^{12}{-}x^{23})(x^{23}{-}x^{31})(x^{31}{-}x^{12})}
{(x^{12}\,x^{23}\,x^{31})^2}\ ,
\end{aligned}
\ee
with \ $R^2=\sfrac13\bigl\{(x^{12})^2+(x^{23})^2+(x^{31})^2\bigr\}$.
A pure Calogero potential is possible only for \ $g=0=\hb$.

\goodbreak

\bigskip

\noindent
{\bf Acknowledgements}

\medskip

\noindent
The author is grateful to Anton Galajinsky and Kirill Polovnikov for
a very fruitful collaboration. His work is partially supported by 
the Deutsche Forschungsgemeinschaft.
\bigskip


\begin{thebibliography}{99}

\bibitem{Wyl99}
N. Wyllard,
J. Math. Phys. {\bf 41} (2000) 2826 [hep-th/9910160].
\bibitem{BGL04}
S. Bellucci, A. Galajinsky, E. Latini,
Phys. Rev. D {\bf 71} (2005) 044023 [hep-th/0411232].
\bibitem{W}
E. Witten,
Nucl. Phys. B {\bf 340} (1990) 281.
\bibitem{DVV}
R. Dijkgraaf, H. Verlinde, E. Verlinde,
Nucl. Phys. B {\bf 352} (1991) 59.
\bibitem{MarGra99}
R. Martini, P.K.H. Gragert,
J. Nonlin. Math. Phys. {\bf 6} (1999) 1 [hep-th/9901166].
\bibitem{Ves99}
A.P. Veselov,
Phys. Lett.  A {\bf 261} (1999) 297 [hep-th/9902142].
\bibitem{ChaVes01}
O.A. Chalykh, A.P. Veselov,
Phys. Lett. A {\bf 285} (2001) 339 [math-ph/0105003].
\bibitem{GLP07}
A. Galajinsky, O. Lechtenfeld, K. Polovnikov,
JHEP 0711 (2007) 008 [arXiv:0708.1075 [hep-th]].
\bibitem{FeiVes07}
M.V. Feigin, A.P. Veselov,
Adv. Math. {\bf 212} (2007) 143 [math-ph/0512095];\\
``On the geometry of $\vee$-systems'',
arXiv:0710.5729 [math-ph].
\bibitem{GLP08}
A. Galajinsky, O. Lechtenfeld, K. Polovnikov,\\
``N=4 mechanics, WDVV equations and roots,''
arXiv:0802.4386 [hep-th].

\end{thebibliography}
\end{document}